# Single-photon source characterization with infrared-sensitive superconducting single-photon detectors

Robert H. Hadfield[a], Martin J. Stevens, Richard P. Mirin, Sae Woo Nam

*National Institute of Standards and Technology, 325 Broadway, Boulder, CO 80305, USA*

(Received…)

Single-photon sources and detectors are key enabling technologies in quantum information processing. Nanowire-based superconducting single-photon detectors (SSPDs) offer single-photon detection from the visible well into the infrared with low dark counts, low jitter and short dead times. We report on the high-fidelity characterization (via antibunching and spontaneous emission lifetime measurements) of a cavity-coupled quantum dot single-photon source at 902 nm using a pair of SSPDs. The twin SSPD scheme reported here is well suited to the characterization of single-photon sources at telecom wavelengths (1310 nm, 1550 nm).

---

[a] Electronic mail: hadfield@boulder.nist.gov

Quantum information processing promises to revolutionize the fields of communications [1] and computing [2] over the coming decades. Many applications in this burgeoning field hinge upon the ability to produce and detect single photons [3]. Practical quantum communications devices will likely take advantage of the existing telecom fiber optic infrastructure, which operates in wavelength bands around 1310 nm and 1550 nm. Thus, the development of more efficient single-photon sources and detectors in the telecom regime is critical to the implementation of quantum information technologies in the real world.

Although some progress has been made in the development of single-photon sources emitting near 1300 nm [4, 5], characterization of these sources has proven quite difficult, owing to the scarcity of good single-photon detectors in this wavelength range. Silicon avalanche photodiodes (APDs), which are the detector of choice for visible-light photon counting [6], are insensitive to wavelengths beyond ~1050 nm. Thus, most measurements for longer wavelengths have relied on InGaAs APDs [7]. These detectors have detection efficiencies (DE) up to 50 % and are typically limited to count rates of ~100 kHz in order to avoid afterpulsing. Furthermore, gating is required to reduce the very high dark count rates.

Nanowire-based superconducting single-photon detectors (SSPDs), which maintain performance at wavelengths beyond 1050 nm, hold great promise for infrared single-photon counting applications. These devices, pioneered by Gol'tsman [8-11] have intrinsic DE up to 20 % in the visible with very few dark counts. They are extremely fast (the minimum reported jitter is 20 ps) and can be clocked at very high rates (up to ~1 GHz). Their single-photon counting capability extends well beyond telecommunications wavelengths (up to 3 μm) [9].

Recently, we characterized a single-photon source emitting at 902 nm by Hanbury Brown-Twiss interferometry coincidence measurements using one silicon APD in tandem with one SSPD [12]. However, questions remained as to whether coincidence measurements could be performed between two or more SSPDs packaged in a single cryocooler without electrical or optical cross talk between the detectors, and whether the DE of the SSPD was adequate to allow the measurement to be carried out with SSPDs alone. In this paper we demonstrate that these technical concerns have been surmounted: we dispense with the Si APD and use a pair of SSPDs—conveniently packaged in a single cryogen-free refrigerator—to successfully characterize a quantum dot-based single-photon source emitting at 902 nm. The data show no indication of cross talk between the two SSPD channels. Since the SSPDs can be operated with very low dark count rates (<10 Hz) even without gating, they make ideal candidates for coincidence measurements of a single photon source pumped by a continuous-wave laser. More importantly, this scheme could be extended to characterize single-photon sources operating beyond 1050 nm.

Each SSPD is based on a narrow superconducting track (100 nm wide, covering a 10 μm x 10 μm area with a 50 % fill factor) of NbN. The photodetection mechanism is as follows: the superconducting track is biased close to its critical current. When the track absorbs a photon, a hotspot is formed, briefly creating a resistive region across the width of the track. The resultant voltage pulse can be amplified and converted into a digital logic pulse. In our detector system, each device is aligned to optical fiber and mounted in a commercial cryogen-free refrigerator, with base temperature of ~2.5 K [12]. The refrigerator has sufficient cooling capacity (100 mW at 4 K) to accommodate multiple fiber-coupled detector channels. In this experiment we use two fiber-coupled SSPDs temperature stabilized at 3.1 K. Our detectors are biased with low-noise current sources with careful attention to grounding to eliminate spurious dark count events induced by electrical pickup. Furthermore we have investigated possible correlations between detection events on the two SSPD channels, both terms of the in the dark counts on the two channels and in terms of photo-induced counts on one channel and dark counts on the other. We have found no evidence of electrical or optical cross talk.

A schematic of the experimental setup is shown in Fig. 1. The single-photon source is an individual InGaAs quantum dot embedded inside a micropillar cavity [13,14]. The optical cavity is formed by a pair of GaAs/AlAs distributed Bragg reflectors grown above and below the dot, and a cylindrical micropillar ~2 μm in diameter is defined with a reactive ion etch. The dot is optically pumped with a Ti:Sapphire laser (~1 ps pulses, center wavelength $\lambda$ = 780 nm, 82 MHz repetition rate) and cooled to ~5 K in a liquid

He flow cryostat. At this temperature, the dot emits single photons at $\lambda$ = 902 nm. The light emitted from the micropillar is collected with an objective lens and directed through a monochromator and into a free-space Hanbury-Brown Twiss (HBT) interferometer. The collection rate of single photons into the interferometer is ~100 kHz. The outputs of the HBT are coupled to telecom fiber and hence to the SSPDs. Fig. 2 shows a measurement of the second-order correlation function of the source, $g^{(2)}(\tau)$, binned at 0.55 ns intervals. The DE of each SSPD detection channel (inclusive of fiber coupling losses from free space into fiber, and from fiber to detector) is only ~2 % at 902 nm, but the dark count rate is extremely low (below 10 Hz per channel). The peaks are spaced at 12 ns intervals (this period corresponds to the laser repetition rate). The width of the coincidence peaks (~1.4 ns FWHM) is dominated by the jitter of the quantum dot single-photon source. Owing to the exceptionally low dark count rates on the detectors, there are very few coincidences between peaks – even though data were acquired over several hours. The area of each peak in the coincidence data is calculated by summing all of the counts in a 3 ns time window. Dividing the area of the peak at $\tau$ = 0 by the mean area of all other recorded peaks yields a second-order intensity correlation $g^{(2)}(0) = 0.081 \pm 0.038$, indicating that the probability of two-photon emission is extremely low. Furthermore, if we subtract the dark counts (spurious coincidences from the counting electronics included), $g^{(2)}(0)$ decreases to ~1 % (although, with uncertainty, this may be as high as 5 %). We have repeated this measurement on the same source under the same conditions using a Si APD in one arm of the HBT interferometer and an SSPD in the other, with a nearly identical result (not shown).

A second important measure of the quality of a single-photon source is the spontaneous emission lifetime, which intrinsically limits the source jitter. The measurement configuration is based on the apparatus depicted in Fig 1. [15]. In this instance a fast photodiode triggered directly by the laser pulse starts the timing electronics, and a single SSPD triggered by the photoluminescence provides the stop. To measure the temporal instrument response function (IRF) of a detector, we tune the monochromator to 780 nm and heavily attenuate the beam, triggering the SSPD directly from the laser. Results are shown in

Fig. 3. The SSPD's response is fit very well—over four decades of dynamic range—by a Gaussian with a FWHM of 68 ps. The Gaussian IRF of the SSPD makes it simple to deconvolve the spontaneous emission lifetime of the source – in this case 400 ps. This setup has been used to perform time-correlated single-photon counting measurements [15] on quantum wells and other semiconductor samples emitting at wavelengths up to 1650 nm.

It should in principle be possible to repeat the $g^{(2)}(\tau)$ experiment described here on a longer-wavelength single-photon source using an identical detection setup – our SSPD system offers 0.5 % DE at 1550 nm with 10 Hz dark count rate per channel [12]. This would undoubtedly slow the rate of data acquisition for a source of equal brightness. Note however that—unlike an InGaAs APD—the SSPD does not need to be gated. Thus one could carry out the $g^{(2)}(\tau)$ measurement with a continuous wave (CW) pump laser, which could lead to a higher collection rate than that for the pulsed pump used here [16]. Since we do not currently have a single photon-source in the 1300 - 1550 nm range available, we have instead used our system to perform a $g^{(2)}(\tau)$ measurement on a pulsed laser at 1550 nm (Fig. 4). The data are binned at 0.55 ns intervals. The laser is gain-switched by voltage pulses from a signal generator at 79 MHz. The mean detected photon number in this measurement is 8.4 x $10^{-4}$. The coincidence histogram is thus proportional to $g^{(2)}(\tau)$ for the source. As expected for a Poissonian source the peak at zero time delay remains and $g^{(2)}(0) = 1$. The breadth of the peaks in the histogram (3.3 ns FWHM) is due to the jitter of the source (2.3 ns).

In conclusion, we have presented the first full characterization of a single-photon source at 902 nm using a pair of superconducting single photon detectors. Our two detector channels are housed within the same cryogen-free refrigerator with no evidence of electrical or optical cross talk. The detection efficiency of our detector system is 2 % per channel, inclusive of coupling losses. However the dark count rate is exceptionally low (10 Hz per channel) compared to APDs. Thus we obtain a $g^{(2)}(\tau)$ measurement of excellent fidelity using twin SSPDs ($g^{(2)}(0) = 0.081 \pm 0.038$). Furthermore we have demonstrated that this detector system can be employed for $g^{(2)}(\tau)$ measurements at 1550 nm, and is thus well suited to the characterization of single-photon sources at telecom wavelengths. The spontaneous emission lifetime measurements on the source highlight the low jitter (68 ps FWHM with Gaussian profile) of this type of detector. With further investment in low-noise electronics we anticipate an improvement in jitter (the lowest reported SSPD jitter is 20 ps FWHM) [11]. Furthermore, new device designs incorporating the detector in an optical cavity may facilitate a considerable boost in available DE in the near future [17].

*Acknowledgements*

This work was carried out with the support of the DARPA QuIST program, the DTO and the U.S. Department of Commerce. We thank N. Bergren and R. Schwall for technical assistance and G. Gol'tsman for providing the detectors used in this work.

*Figure Captions*

Figure 1: Setup for $g^{(2)}(\tau)$ experiment. A standard Hanbury-Brown Twiss configuration is used with SSPDs at each output. In the lifetime measurement a fast photodiode is instead of one of the SSPDs as a start trigger for the timing electronics.

Figure 2: $g^{(2)}(\tau)$ for a cavity-coupled quantum dot using twin SSPDs. The SSPDs have a detection efficiency (including fiber coupling losses) of 2 % per channel. They are biased such that the dark count rate is ~ 10 Hz. $g^{(2)}(0) = 0.081 \pm 0.038$.

Figure 3: Spontaneous emission lifetime of a quantum dot measured with a single SSPD channel. The start clock to timing electronics is provided by a fast photodiode driven directly by the Ti: Sapphire pump laser. The carrier lifetime of the source is ~ 400 ps. The instrument response function (IRF) of the SSPD is a 68 ps FWHM Gaussian.

Figure 4: $g^{(2)}(\tau)$ of a 1550 nm diode laser gain switched at 1550 nm measured with twin SSPDs. As expected for a Poissonian source, $g^{(2)}(0) = 1$. The width (3.3 ns FWHM) of the peaks is dominated by the jitter (2.3 ns FWHM) of the laser.

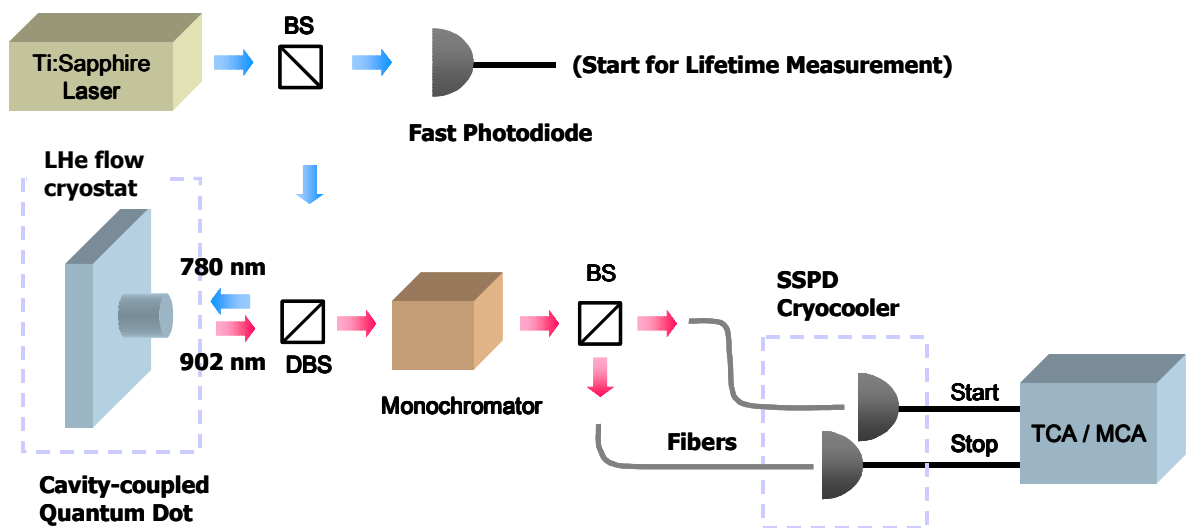

Figure 1

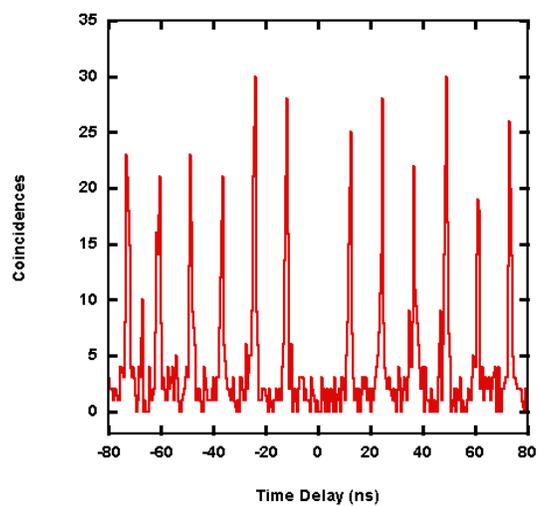

Figure 2

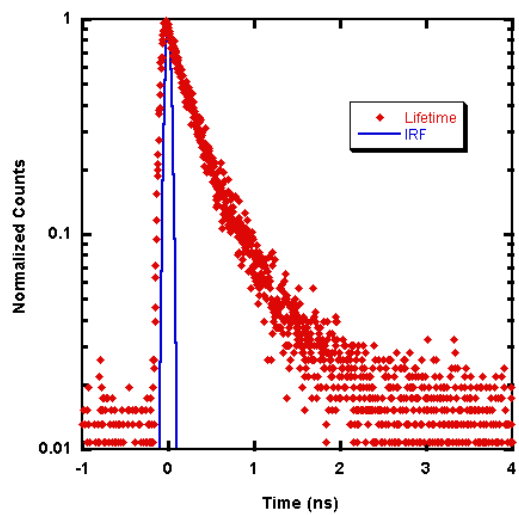

Figure 3

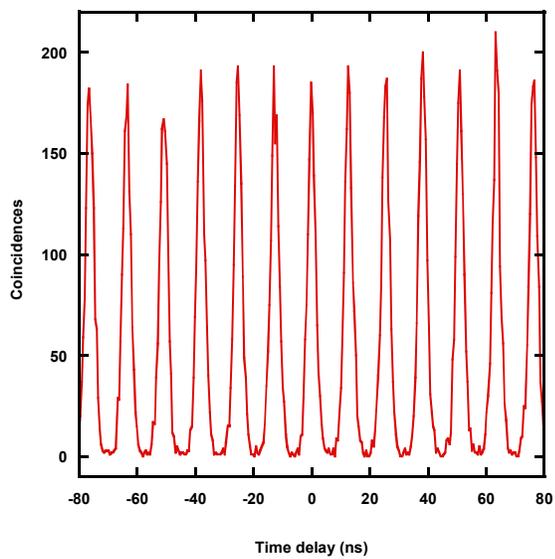

Figure 4